\documentclass[pra,superscriptaddress,twocolumn]{revtex4-1}[11pt] 
\pdfoutput=1
\usepackage{amsmath,amssymb,graphicx} 
\usepackage{hyperref}
\usepackage{color}

\begin{document}

\title{Electron train backgrounds in liquid xenon dark matter search detectors are indeed due to thermalization and trapping}
\author{P. Sorensen}
\thanks{pfsorensen@lbl.gov}
\affiliation{Lawrence Berkeley National Laboratory, 1 Cyclotron Rd., Berkeley, CA 94720, USA} 
 
\begin{abstract}
Electron emission from liquid into gaseous xenon is a cornerstone of dark matter search detectors such as ZEPLIN, XENON, LUX and LZ. The probability of emission is a function of the applied electric field $E$, and electrons which fail to pass from the liquid into the gas have been previously hypothesized to become thermalized and trapped. This article shows, for the first time, quantitative agreement between an electron emission model and existing data.  The model predicts that electrons in the liquid must surmount a typical potential barrier $\phi_b=0.34\pm0.01$~eV in order to escape into the gas. This value is a factor of about $\times2$ smaller than has previously been calculated or inferred. Knowledge of $\phi_b$ allows calculation of the lifetime of thermalized, trapped electrons. The value is $\mathcal{O}(10)$~ms, which appears to be compatible with XENON10 observations of electron train backgrounds. As these backgrounds limit the sensitivity of dark sector dark matter searches, possible mitigations are discussed.
\\
\end{abstract}
\date{\today}

\maketitle


\section{{Introduction}}\label{sec:intro}
Particle detection with low-background liquid/gas xenon detectors offer world-leading sensitivity to hypothetical dark matter particle scattering events \cite{Akerib:2016vxi,Aprile:2012nq}. In the absence of any convincing evidence for detection, the experimental challenge remains to increase detector sensitivity. This primarily means larger target masses and lower backgrounds. It  also means maximizing the search capabilities of each particular detector. An example of this is the so-called ``S2-only'' or electron-only search technique \cite{Angle:2011th}. The technique is sensitive to single electrons, due to the fact that ionized electrons are extracted into the gas phase and amplified via proportional scintillation. With sensitivity set by the energy threshold of a single quanta, liquid xenon targets are able to search for lower ($\lesssim10$~GeV) mass WIMP dark matter candidates, as well as for dark sector candidates \cite{Essig:2011nj,Essig:2012yx}. 

However, ``electron train'' backgrounds present a serious limitation to this technique. These consist of numerous single electron signals, typically of width $\sim~1~\mu$s, emitted over timescales $\gg 100~\mu$s. The time profile of this background is therefore sparse and long-lived compared with typical event durations, which are $\mathcal{O}(100)~\mu$s. It was suggested in Ref. \cite{Angle:2011th} that the origin of electron train backgrounds might lie in the thermalization and trapping of un-emitted electrons just below the liquid xenon surface, with eventual emission on significantly longer timescales. This possibility was previously hypothesized in Ref. \cite{Bolozdynya:1999a}. However, the hypothesis remains unconfirmed and the trapping lifetime is unknown.  As a result, the dark matter search community sentiment appears to have concluded that liquid xenon will not be a competitive target for observing single or few electron signals from putative dark sector dark matter \cite{Alexander:2016aln} (see in particular Fig. 12 and Fig. 13, where a projection for xenon is omitted entirely). This sentiment was likely reinforced by a recent S2-only analysis \cite{Aprile:2016wwo}, which stated that {\it ``a full background model ...  cannot be constructed, as the origin of the small-S2 background in the detector cannot be reliably quantified...''}

The purpose of this article is to show that the thermalized, trapped electron hypothesis appears to be correct and that the background can in fact be quantified.

\section{{Preliminaries}}\label{sec:refresh}
We remind the reader that the standard search technique with this class of detector requires putative dark matter scattering events to result in both scintillation photon (S1) and ionized electron (S2) signals, separated by the time $\Delta t$ required for electrons to drift a distance $\Delta z$ across the liquid xenon target. The maximum drift time of an electron from the cathode to the liquid surface is $\Delta t_{max}$. For post-event time windows $<\Delta t_{max}$, photoionization is the primary cause of single electrons \cite{Edwards:2007nj}. Photoionization is easily vetoed in an S2-only analysis, because $\Delta t_{max}$ is typically a few hundred $\mu$s and so is directly connected to the originating event.  In contrast, electron train backgrounds are observed to continue for time periods $\gg \Delta t_{max}$.  In order to calculate the lifetime of trapped electrons, it is necessary to know the barrier height imposed by the dielectric interface between the liquid and the gas phase xenon. This same barrier governs the emission of electrons which have been heated by the applied electric field. The first step is therefore to study the electron emission efficiency. 

\section{Electron emission efficiency}\label{sec:eee}
Ionized electrons in liquid xenon may be readily drifted by application of an electric field $E_d$, with typical values $E_d\sim0.2-0.7$~kV/cm. Such electrons are referred to as quasi-free due to being confined to the medium, and yet not localized to any particular atom. Dual phase (liquid/gas) xenon detectors for dark matter search have a Frisch grid a few mm below the liquid surface and an anode grid a few mm above it. A much larger electric field $E$ may then be applied across the phase boundary. Typical ranges of the applied electric fields in the liquid are $E\sim3-5$~kV/cm. The electric field in the gas phase just above the liquid surface is approximately $2E$ due to the relative dielectric constant of liquid xenon, $\epsilon=1.96$ \cite{Sawada:2003a}. In this work, applied electric fields are assumed to have only a $z$ component.  A temperature T=173~K was assumed for all calculations. Typical operating temperatures tend to be within $\pm10$~K of this value, and represent a rather small effect.

\begin{figure}[h]
\begin{center}
\vskip -0.0cm
\includegraphics[width=0.45\textwidth]{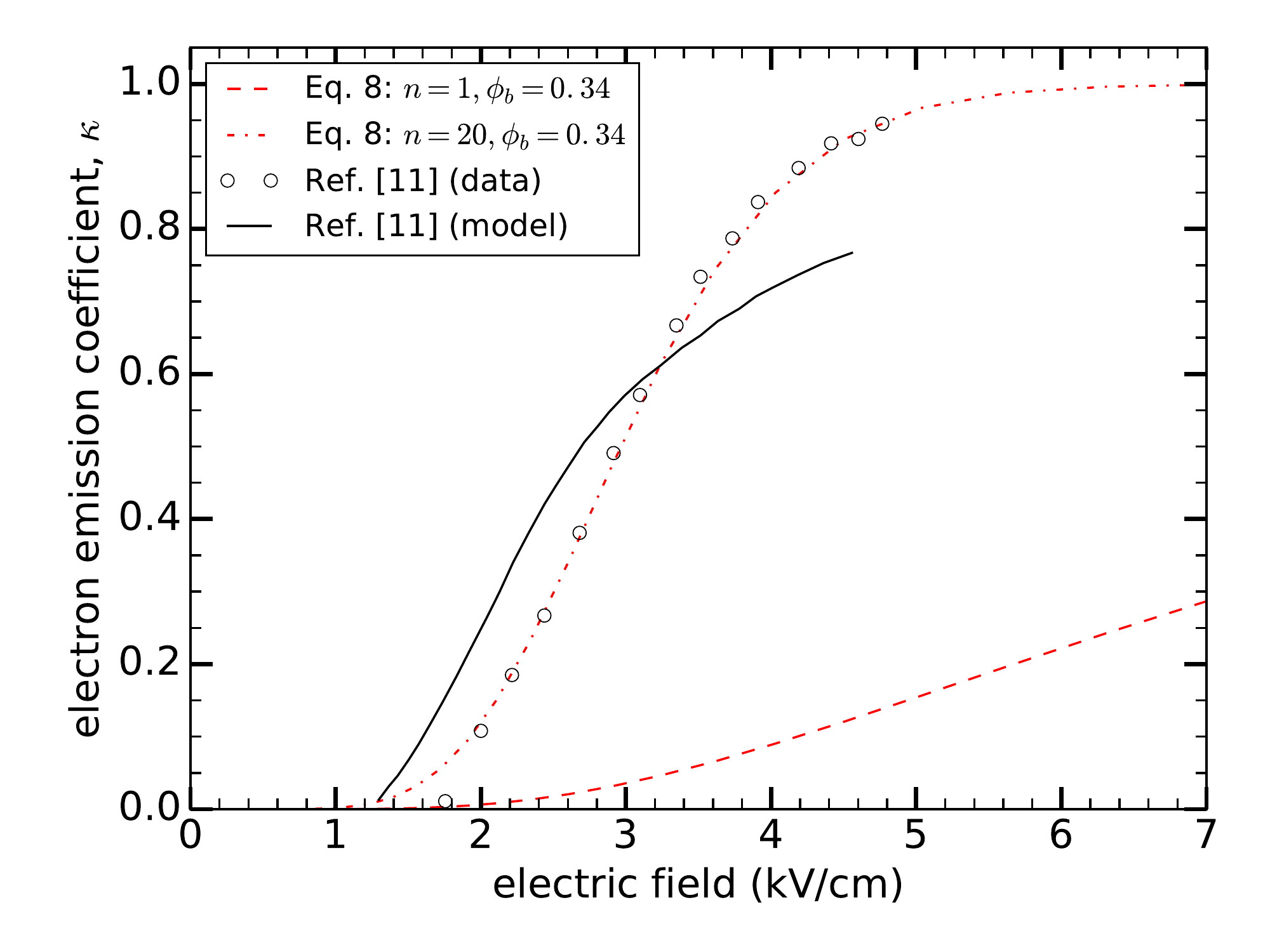}
\vskip -0.1cm
\caption{Absolute efficiency for electron emission from liquid into gas xenon, as a function of the electric field in the liquid phase.  }
\vskip -0.5cm
\label{fig:emission}
\end{center}
\end{figure}

The emission of quasi-free electrons across the liquid-gas boundary in pure xenon is a strong function of the applied electric field. Only a single absolute measurement exists \cite{Gushchin:1982a}, which is reproduced in Fig. \ref{fig:emission}. Several relative measurements (see e.g. \cite{Aprile:2013blg,Bernard:2016uc}) agree quite well with that work if they assume that at the highest measured values of $E$, the efficiency for emission is unity. Because of this extra assumption, relative measurements are not considered here.

The authors of Ref. \cite{Gushchin:1982a} suggested a Schottky barrier model for the electron emission efficiency, and their best fit is reproduced in Fig. \ref{fig:emission}. The agreement is not particularly satisfying, a point perhaps noted in the closing statement of that work: {\it``it appears that the mechanism of electron emission from liquid noble gases is more complicated ...''}. It is notable that both data and model from Ref. \cite{Gushchin:1982a} suggest the possibility that the high-$E$ limit of the emission efficiency might not be unity. 

\subsection{The Shottky barrier model}
\begin{figure}[h]
\begin{center}
\vskip -0.0cm
\includegraphics[width=0.45\textwidth]{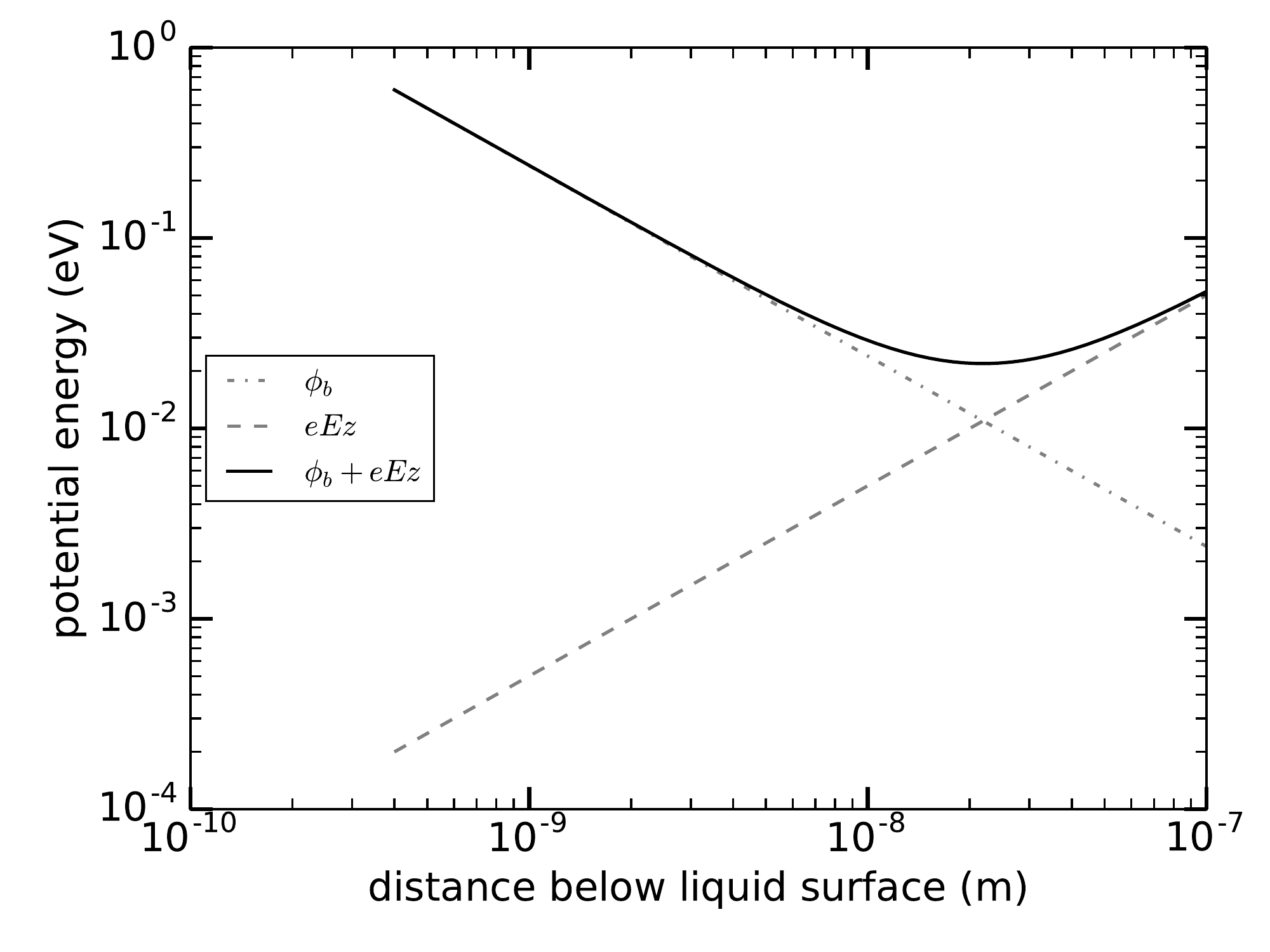}
\vskip -0.1cm
\caption{Potential energy of an electron just below the liquid-gas interface.  }
\vskip -0.5cm
\label{fig:potential}
\end{center}
\end{figure}

As an electron approaches a dielectric boundary that is held at a constant potential, the force due to it's image charge results in an energy barrier. This is sometimes referred to as the Schottky barrier, given by
\begin{equation}\label{eq:phib}
\phi_b = \frac{e^2}{8\pi \epsilon_0 z}\frac{\epsilon-1}{\epsilon+1}.
\end{equation}
The force driving the electron toward the barrier is simply that due to the applied electric field, so $\phi=e E z$. The combined potential is shown in Fig. \ref{fig:potential}. For electrons in the condensed state, collisions tend to randomize an electron's velocity. Therefore Fig. \ref{fig:potential} is only valid within a single scattering length of a drifting electron.

The external field $E$ does two things with respect to electron emission: it increases the energy of the drifting electrons, and it lowers the height of the barrier  by an amount equal to
\begin{equation}\label{eq:deltaphib}
\Delta\phi_b = e \left(\frac{eE}{4\pi \epsilon_0 z}\frac{\epsilon-1}{\epsilon+1}\right)^{1/2}
\end{equation}
Equations \ref{eq:phib} and \ref{eq:deltaphib} can be readily derived from electrostatics. The energy distribution $f_0(\varepsilon)$ of electrons drifting under the influence of an electric field is less obvious. It may be derived from a solution of the Boltzman equation, as explained in {\it ``Theory of hot electrons in liquids, gases and solids.''} \cite{Cohen:1967a}. One finds that on average, electrons retain thermal energies for electric fields below about 50~V/cm. By 5~kV/cm, the average energy of electrons is a factor $\times10$ above thermal.

From these considerations, the authors of Ref. \cite{Gushchin:1982a} define the electron emission efficiency as
\begin{equation}\label{eq:emis}
\kappa = \int_{\phi_b-\Delta\phi_b}^\infty \varepsilon^{1/2} f_0(\varepsilon) d\varepsilon \bigg/  \int^\infty_0 \varepsilon^{1/2} f_0(\varepsilon) d\varepsilon,
\end{equation}
in which the factor $\varepsilon^{1/2}$ serves to select electrons whose velocity has a component directed toward the barrier.  Equation \ref{eq:phib} tends to infinity as the electron approaches the liquid/gas interface. Physically, this is accommodated by the fact that the surface thickness is defined by atoms. If one assumes liquid xenon atoms to be arranged as a simple cubic lattice, the lattice constant can be calculated to be about $4\times10^{-10}$~m. Inserting this into Eq. \ref{eq:phib} results in $\phi_b = 0.61$~eV, which is the value quoted in Ref. \cite{Bolozdynya:1999a}. In Ref. \cite{Gushchin:1982a}, $\phi_b$ was treated as a free parameter and was determined to have the value $\phi_b = 0.84$~eV. The present work also treats $\phi_b$ as a free parameter.


There are two problems with this basic model. The first concerns the mean free path, and the second concerns the number of attempts an electron makes before it succeeds in crossing the barrier. 

\subsection{The n$^{th}$ chance model}
The n$^{th}$ chance model begins with the same premises as the Shottky barrier model. As a preamble, it is critical to correct the values of the mean free paths which underpin the electron energy distribution $f_0(\varepsilon)$. Two mean free paths are defined in Ref. \cite{Cohen:1967a}. In Ref. \cite{Gushchin:1982a} these were assumed to have the values $\Lambda_0 \sim 10^{-9}$~m and $\Lambda_1 \sim 10^{-7}$~m.  However, calculations suggest $\Lambda_0 = 3.6\times10^{-9}$~m and $\Lambda_1$ is a function of $E$ \cite{Sowada:1975a}. The field dependence given in Ref \cite{Sowada:1975a} can be parameterized as
\begin{equation}\label{eq:lam}
\Lambda_1 = \frac{\sqrt{2} \times 10^{-7}}{1+(0.91E\times10^{-4})^{2/3}}.
\end{equation}
It is easy to verify that this parameterization is reasonable by calculating the electron drift velocity within the same framework \cite{Cohen:1967a}. It is given by
\begin{equation}\label{}
v_d = \frac{1}{3}(\frac{2}{m})^{1/2} \left( \int_0^\infty \varepsilon (-E \Lambda_1 \frac{df_0}{d\varepsilon}) d\varepsilon  \bigg/  \int_0^\infty \varepsilon^{1/2} f_0 d\varepsilon \right)
\end{equation}
in which 
\begin{equation}\label{eq:f0}
f_0 = (\varepsilon/kT + b)^b e^{-\varepsilon/kT}
\end{equation}
and
\begin{equation}\label{eq:b}
b = \frac{1}{3}(eE\Lambda_0)(eE\Lambda_1) / (2m/M)(kT)^2.
\end{equation}

\begin{figure}[h]
\begin{center}
\vskip -0.0cm
\includegraphics[width=0.45\textwidth]{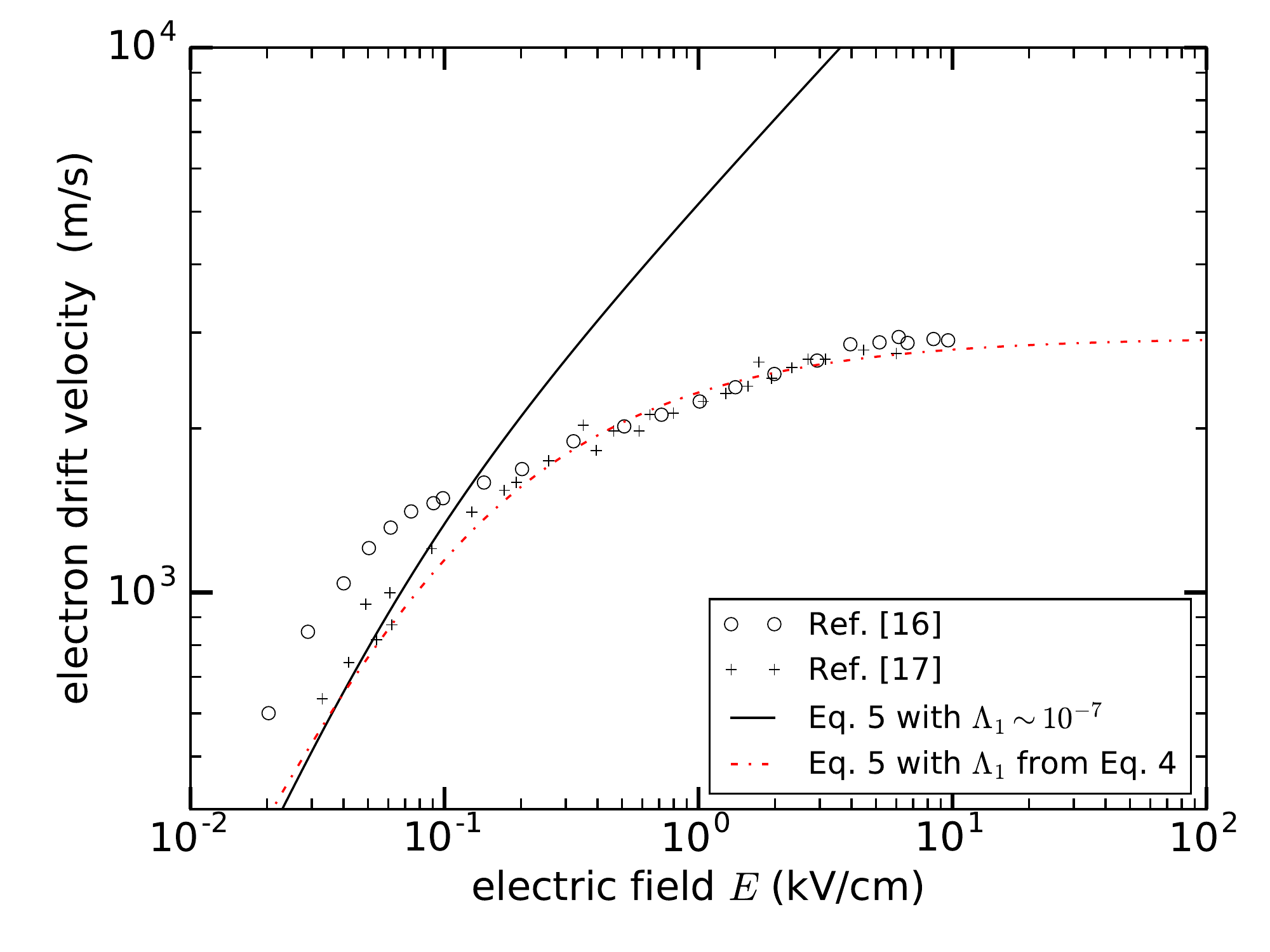}
\vskip -0.1cm
\caption{ Measured electron drift velocities in liquid xenon are compared with calculated values. }
\vskip -0.5cm
\label{fig:mobility}
\end{center}
\end{figure} 

In the previous equation, $m$ and $M$ are the masses of an electron and an atom, respectively. It is then apparent from Fig. \ref{fig:mobility} that in Ref. \cite{Gushchin:1982a}, the electron energy distribution $f_0$ was incorrectly biased towards unrealistically large electron energies. With $\Lambda_1$ corrected by Eq. \ref{eq:lam}, it is impossible to reproduce the theoretical curve (labeled ``Ref. [10] model'' in Fig. \ref{fig:emission}), for any choice of $\phi_b$.

This brings us to the second problem with the simple Shottky barrier model: electrons scatter significantly as they travel through matter, their trajectory a random walk with a slight bias provided by the applied electric field. We must suppose that an electron which fails to cross the barrier on it's initial attempt will most likely continue to scatter. It may yet escape into the gas, or it may eventually thermalize just below the surface in the potential minimum shown in Fig. \ref{fig:potential}. This scenario was discussed in Ref. \cite{Gushchin:1982a}, but evidently not implemented in their model.

\begin{figure}[h]
\begin{center}
\vskip -0.0cm
\includegraphics[width=0.45\textwidth]{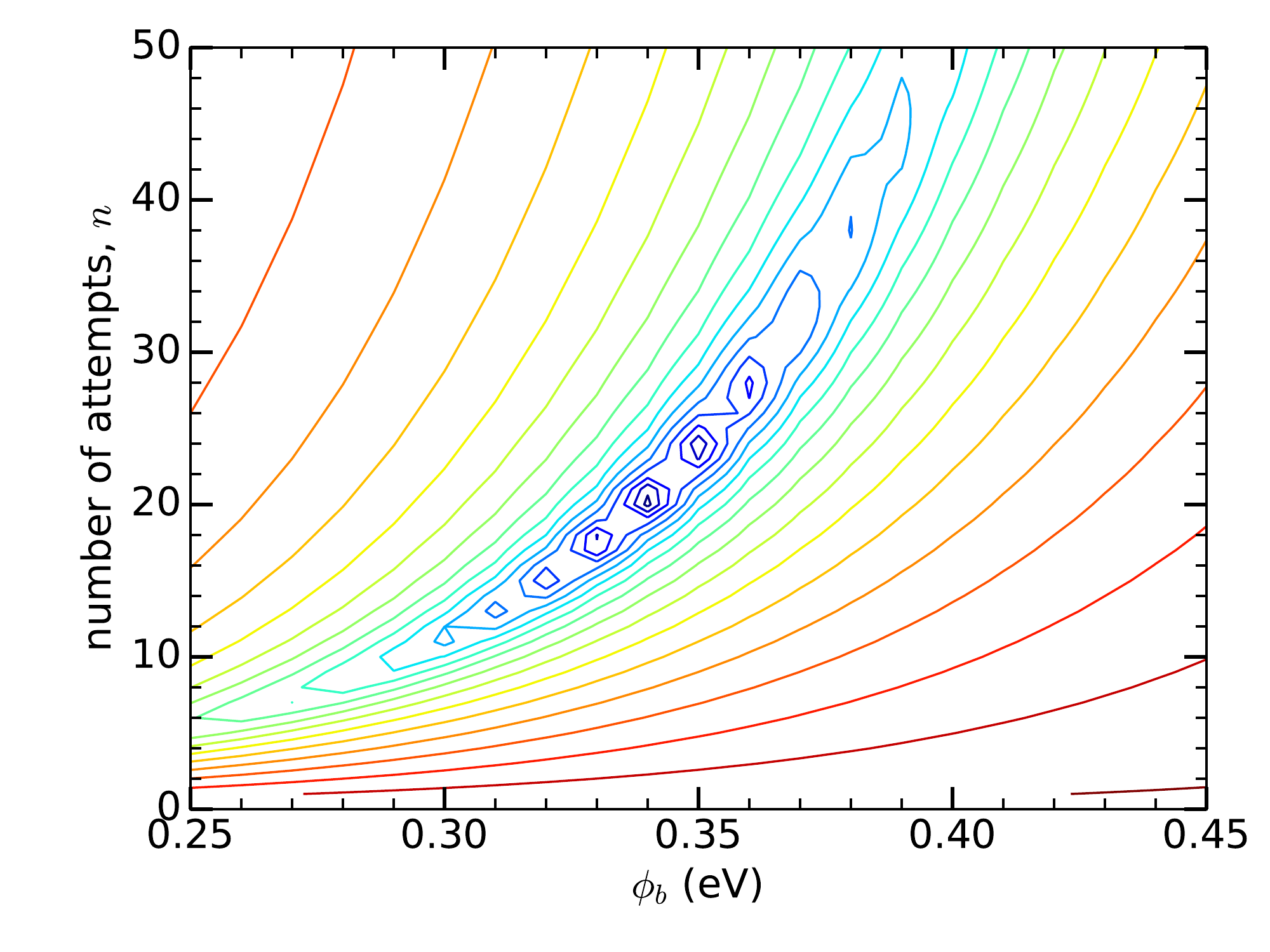}
\vskip -0.1cm
\caption{Contour plot of $\chi^2$ for comparison of the emission model given by Eq. \ref{eq:nth} vs the data shown in Fig. \ref{fig:emission}. Agreement at $2\sigma$ is confined to $\phi_b=0.34\pm0.1$~eV.  }
\vskip -0.5cm
\label{fig:heat}
\end{center}
\end{figure}

Since Eq. \ref{eq:emis} gives the probability of electron emission for a single attempt, the cumulative probability of emission after $n$ attempts is simply 
\begin{equation}\label{eq:nth}
\kappa_n = 1-(1-\kappa)^n.
\end{equation}
This model was compared with the data over the range $1\leq n \leq 100$ and $0.05 \leq \phi_b \leq 1.0$~eV. The relevant region is shown in Fig. \ref{fig:heat}. A best fit is found for $\phi_b = 0.34\pm0.01$~eV, corresponding to approximately $n=20\pm4$. This result is not too dissimilar from the value $\phi_b=0.38$ that is obtained from Eq. \ref{eq:phib} for $z=6.2\times10^{-10}$~m, the measured lattice constant of solid xenon \cite{Wyckoff:1963}.

The best fit emission efficiency curve is also plotted in Fig. \ref{fig:emission}. It should be noted that the $n$ appearing in this model is really an expectation value $\langle n \rangle$. The variation in $n$ is not addressed here.

At an electric field $E=5$~kV/cm, the mean free path between collisions is $\Lambda_1\approx10$~nm and the saturated drift velocity is nearly 3000~m/s. This suggests a typical relaxation time $\tau$ between collisions of about 3~ps, and sets the timescale for emission in the $n^{th}$~chance model. Suppose that a typical electron experiences up to 10 additional collisions between successive attempts to escape the liquid. The emission process would be complete in less than 1~ns, which is consistent with observation.

\section{Lifetime of thermalized, trapped electrons}\label{sec:et}
\begin{figure}[h]
\begin{center}
\vskip -0.0cm
\includegraphics[width=0.45\textwidth]{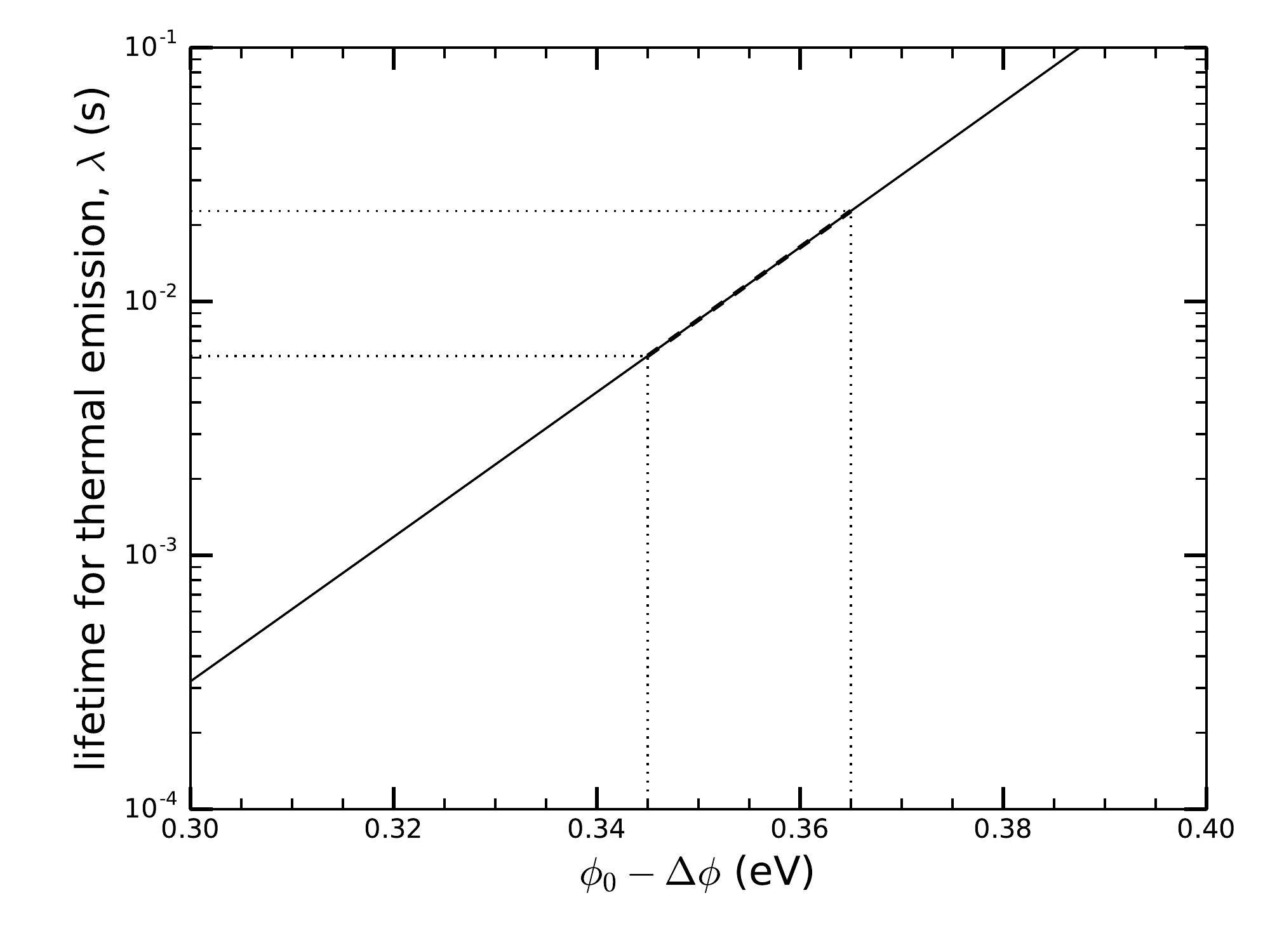}
\vskip -0.1cm
\caption{The lifetime for emission of thermalized electrons which are trapped below the liquid surface, assuming $\Delta \phi_b = 0.015$~eV. This value is appropriate for $E=5$~kV/cm.  }
\vskip -0.5cm
\label{fig:etrain}
\end{center}
\end{figure}

The results of the preceding section allow calculation of the lifetime of thermalized, trapped electrons. It is given by
\begin{equation}
\lambda = \tau / \kappa_{b}.
\end{equation}
Here $\kappa_b$ is calculated from Eq. \ref{eq:emis}, with $f_0=e^{-\varepsilon/kT}$ as the limiting case of Eq. \ref{eq:f0} when $E\rightarrow0$. Assuming a relaxation time $\tau=3$~ps and taking the result $\phi_b=0.34\pm1$~eV, Fig. \ref{fig:etrain} predicts a lifetime of trapped electrons in the range $6-23$~ms. The prediction appears to be compatible with the original observations of XENON10. There is some uncertainty about the correct relaxation time for this calculation: trapped electrons are in a region where the $z$ component of the electric field reverses direction. It seems possible that the relaxation time for trapped electrons could be larger by an order of magnitude or more. This would directly increase the calculated lifetime for emission.

Thermal electrons may undergo diffusion prior to emission. The characteristic scale of this diffusion is $\sigma~=~\sqrt{2Dt}$. Taking $D=80$~cm$^2$/s \cite{Doke:1981} and $t=10$~ms as typical values results in $\sigma=1.3$~cm. This suggests that electron train backgrounds may be emitted many cm from the $(x,y)$ vertex of the originating event.

It must be noted that this simple model has ignored the inevitable complications that might be encountered in a real detector, such as surface roughness (waves), a radial component to the electric field $E$, a periodic potential due to the electrode wires or meshes, and fluid flow, evaporation and condensation. Further investigation by experimental groups is essential, and would ideally compare a measured thermal emission lifetime against this model. Such measurement appears to require low-background conditions.



\section{Possible Mitigation}
There is significant interest in suppressing the electron train background, as it is an impediment to searches for low-mass WIMP dark matter or dark sector dark matter candidates. The calculated lifetime is very long and suggests that this background is inherent to the operation of liquid/gas xenon detectors. However, there may be several ways to suppress the background. The first possibility is of course to increase the emission probability $\kappa$. The model suggests $\kappa\simeq99.9\%$ at a field $E\simeq7$~kV/cm. This value is only slightly larger than the operating fields of previous and existing experiments, but may be challenging to achieve. However, it would immediately offer a factor of about $\times50$ suppression.

A more elaborate suppression of the background could in principle be implemented with fast ($\sim \mu$s) high voltage switching \cite{Shutt:pc}, such that immediately after registering the S2 signal from an event, trapped electrons from that event would be directed back to the electrode located a few mm below the liquid surface. Alternatively, it may be possible to stimulate the emission of the electrons using LED photons. For example, commercially available 940~nm LEDs would provide a typical energy of 1.3~eV per photon, more than enough to help trapped electrons surmount $\phi_b$. The photocathode sensitivity of the photomultipliers used for detection of xenon scintillation light typically decrease by over three orders of magnitude by 700 nm. Likely, they would be blind to these photons.

A final comment is that should all hardware mitigation efforts fail, it should still be possible to mitigate electron train backgrounds in off-line analysis. By looking at tens of ms windows following large $\gtrsim1$~MeV events, the time structure ($\lambda$) of single electron emission could be measured. Then, an $(x,y)$ map of the variation in $\lambda$ could be constructed. Such a map would likely be necessitated by experimental complications, an incomplete list of which can be found at the end of Sec. \ref{sec:et}. Finally, a background expectation could be placed on all inter-event quiet periods in the detector, regardless of the size of the preceding event. This could significantly increase the sensitivity of future searches for low mass or dark sector dark matter candidates. 


\clearpage{\pagestyle{empty}\cleardoublepage}
\section*{Appendix}
Liquid xenon detectors deployed for particle physics experiments have a track record of operating at cathode voltages which are significantly lower than their design goal. The operating voltage in some cases has been limited by the onset of unexplained, voltage-dependent scintillation. This scintillation can quickly overwhelm any potential signal, and thus operation of the detector. Could it be that the effect originates from electrons drifting in very small regions of very high electric field? What follows is a calculation of this possibility.


The the probability to encounter an electron with energy $\varepsilon$ greater than some critical value $\varepsilon_c$ is
\begin{equation}\label{eq:tails}
p = \int_{\varepsilon_c}^\infty \varepsilon f_0(\epsilon) d\varepsilon \bigg/  \int^\infty_0 \varepsilon f_0(\varepsilon) d\varepsilon.
\end{equation}

\begin{figure}[h]
\begin{center}
\vskip -0.0cm
\includegraphics[width=0.45\textwidth]{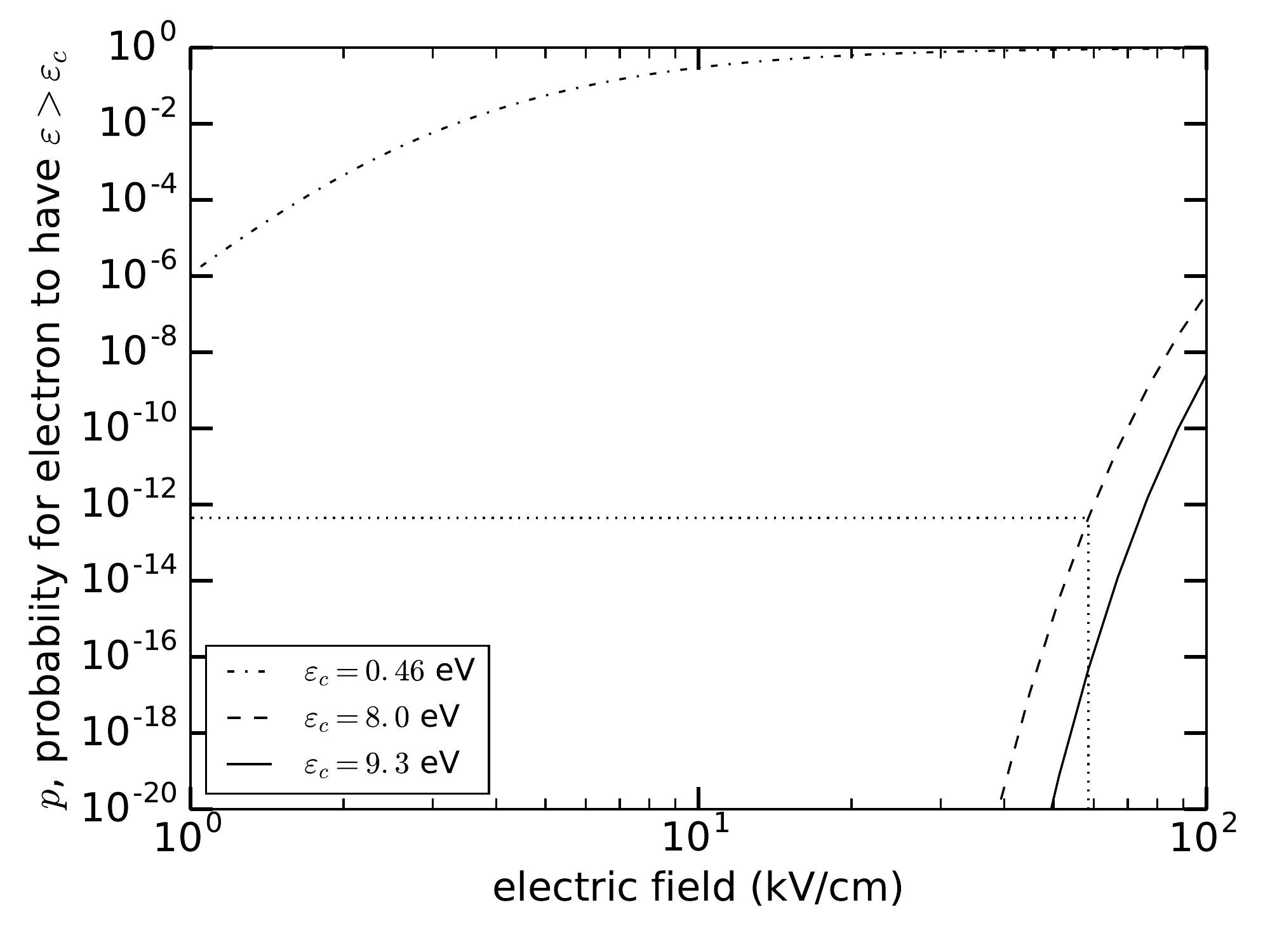}
\vskip -0.1cm
\caption{The probability for a drifting electron to acquire an energy $\varepsilon > \varepsilon_c$. These cases are described in the text.  }
\vskip -0.5cm
\label{fig:tails}
\end{center}
\end{figure}

The behavior of Eq. \ref{eq:tails} is plotted in Fig. \ref{fig:tails} for three choices of $\varepsilon_c$. The first corresponds to the electron affinity of O$_2^-$ ($\varepsilon_c=0.46$~eV), a common electronegative impurity in liquid xenon.  The other two values are related to the xenon itself. In gas, the energy of the excited state Xe$^*$ is $\varepsilon=8.4$~eV and the first ionization potential is 12.1~eV \cite{Saloman:2004}. In the liquid state, the band gap is about 25\% smaller, at 9.3~eV \cite{Asaf:1974}. The energy of Xe$^*$ in the liquid state also appears to be smaller, and also to broaden somewhat, with it's lower bound at about 8.0~eV \cite{Beaglehole:1965}.

Suppose a 1~mm region of liquid xenon is subject to an electric field $E=60$~kV/cm. The number of excited xenon atoms that might be produced by a single electron traversing this region would be given by $N = (\tau v_d / p) \times10^{-3}$. The dotted lines in Fig. \ref{fig:tails} point to the value $p\approx10^{-12}$ in this case, so one finds $N\approx9$. Each excited xenon atom  is understood to lead to the emission of a single scintillation photon via the process 
\begin{equation}
\mbox{Xe}^* + \mbox{Xe} \rightarrow \mbox{Xe}_2^* \rightarrow 2\mbox{Xe} + h\nu
\end{equation}

It must be noted that the analysis in this section relies on an extrapolation of $\Lambda_1$, as indicated in Fig. \ref{fig:mobility}. While the extrapolation appears reasonable, it introduces additional uncertainty of at least a factor of $\times2$. It is also worth noting that the present result suggests an onset of scintillation at a value of $E$ which is an order of magnitude smaller than that found in Ref. \cite{Aprile:2014ila}.






\end{document}